# Thermodynamic properties of a diluted triangular Ising antiferromagnet in a field

M. Borovský, M. Žukovič*, A. Bobák

Institute of Physics, Faculty of Sciences, P. J. Šafárik University, Park Angelinum 9, 041 54 Košice, Slovakia

Within the framework of the effective-field theory with correlations we investigate effects of an external magnetic field and random site dilution on basic thermodynamic quantities, such as the magnetization and the magnetic susceptibility, on the geometrically frustrated triangular lattice Ising antiferromagnet. Behavior of these quantities is presented in the temperature-field parameter space for selected mild degrees of dilution. It is found that, besides the anomalies associated with phase transitions from the ferrimagnetic to the paramagnetic state, in certain regions of the parameter space these functions display some more anomalies and peculiarities, as a result of joint effects of the geometrical frustration, magnetic dilution, thermal fluctuations and the applied magnetic field.

PACS numbers: 05.50.+q, 64.60.De, 75.10.Hk, 75.50.Ee

### 1. Introduction

According to known exact results, geometrical frustration in the Ising antiferromagnet on the triangular lattice is responsible for lack of long-range ordering at any temperature [1]. This situation can be changed by lifting degeneracy from the system with either applied magnetic field or injection of quenched nonmagnetic impurities. In the former case, we can observe a broad 1/3 plateau in the field dependency of the magnetization at low temperatures due to ferrimagnetic ordering, when two sublattices are oriented parallel and one opposite to the field direction (↑↑↓) [2]. Such a behavior has already been observed in experimental measurements of some frustrated antiferromagnetic compounds, such as $Ca_3Co_2O_6$ [3]. In the case of the random dilution with nonmagnetic impurities, the frustration is relieved only locally, which can lead to the spin-glass state [4]. Furthermore, the dilution of the system in the presence of the field shatters the 1/3 magnetization plateau into step-like curve [5,6]. This interesting phenomenon is expected to also reflect in the behavior of the magnetic susceptibility.

The goal of the present paper is to explore the magnetization and the magnetic susceptibility as continuous functions in the temperature-field parameter plane and their evolution with increasing dilution.

### 2. Effective-field theory

The randomly diluted Ising model in the external magnetic field is described by the Hamiltonian

$$H = -J\sum_{\langle i,j \rangle} \xi_i \xi_j S_i S_j - h \sum_i \xi_i S_i , \quad (1)$$

where $S_i = \pm 1$, $\langle i,j \rangle$ denotes the sum over all nearest neighbor pairs, $J < 0$ is the antiferromagnetic interaction constant, $h$ is the external magnetic field, and $\xi_i$ is a random variable which is either 1, when site $i$ is occupied with magnetic atom, or 0 otherwise. The configurational average of $\xi_i$ represents the concentration of magnetic atoms $p$.

The total magnetization of the system $m$ is calculated as a mean value of the sublattice magnetizations, i.e., $m = (m_A + m_B + m_C)/3$. Within the effective-field theory with correlations [6], we can obtain a set of coupled nonlinear equations (see Ref. [6] for details)

$$\begin{aligned} m_A &= p(a+bm_B)^3(a+bm_C)^3 \tanh(x+\beta h)|_{x=0}, \\ m_B &= p(a+bm_A)^3(a+bm_C)^3 \tanh(x+\beta h)|_{x=0}, \\ m_C &= p(a+bm_A)^3(a+bm_B)^3 \tanh(x+\beta h)|_{x=0}, \end{aligned} \quad (2)$$

where $a = 1 - p + p\cosh(\beta JD)$, $b = \sinh(\beta JD)$, $\beta = 1/k_BT$, and $D = \partial/\partial x$ is the differential operator.

The magnetic susceptibility is defined as $\chi = \partial m/\partial h|_{T=const.}$, and can be calculated as $\chi = (\chi_A + \chi_B + \chi_C)/3$, where $\chi_i$, $i = A, B$ or $C$, are sublattice susceptibilities. The latter can be obtained by differentiating equations (2) and subsequent solving of a set of the resulting linear equations for $\chi_A$, $\chi_B$ and $\chi_C$. Alternatively, $\chi$ can be obtained as a numerical derivative of $m$ with respect to $h$ at constant $T$.

### 3. Results and discussion

In Fig. 1 we present the results for the pure (p=1) system. The low-temperature 1/3 magnetization plateau in the ferrimagnetic (↑↑↓) phase within $0 < h/|J| < 6$ is clearly visible in the contour plot in Fig. 1(a). As the temperature is increased, the plateau gradually dissolves and above $k_BT/|J| \approx 1.35$ it vanishes completely. The onset of the magnetization plateau is accompanied with the inverted U-shape curve of the susceptibility peaks (the thick rugged curve shown in Fig. 1(b)). The heights of the peaks decrease with the increasing temperature. Further, we can notice that within some rage of sufficiently low temperatures, the curve of the susceptibility peaks forks out around the field values $h/|J| = 0$ and 6. Hence, roughly within $0 < k_BT/|J| < 0.2$, are two more peaks which result from inflection points in the $m$-$h/|J|$ curves near $h/|J| = 0$ and 6, observed also in the Monte Carlo simulations [5].

*corresponding author; e-mail: milan.zukovic@upjs.sk

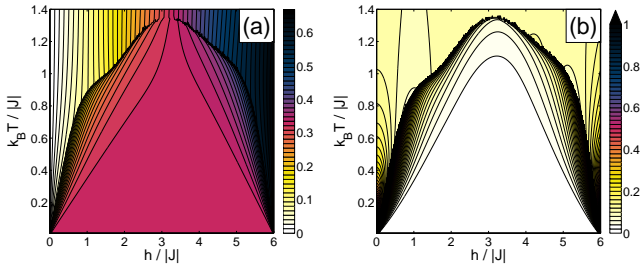

Fig.1. Filled contour plots of the field-temperature dependencies of (a) the total magnetization $m$ and (b) the susceptibility $\chi$ for $p = 1.0$.

In the following, we will be interested in how the presence of small amount of magnetic impurities can change this behavior. In particular, we will first consider the concentration of magnetic atoms $p = 0.95$. In accordance with the previous studies [5,6], the low-temperature 1/3 magnetization plateau starts breaking up into smaller steps at about integer values of the field (Fig. 2(a)). When we gradually increase the temperature, the steps become smoother until the paramagnetic phase is reached and the magnetization-field dependence becomes linear above $k_BT/|J| \approx 1.1$. By checking the sublattice magnetizations we can find out that close to zero temperature the ferrimagnetic phase does no longer extend from $h_{min} = 0$ to $h_{max} = 6$ but only to $h_{max} = 5$. Nevertheless, at higher temperatures it can persist even a bit above $h_{max} = 5$, which leads to the reentrant phenomenon. This situation is reflected in the susceptibility dependence, shown in Fig. 2(b). Namely, the inverted U-shape curve of the maxima associated with the phase transition is deformed to connect the values $h/|J| = 0$ and 5 and the temperature dependence at a fixed value of the field just above $h/|J| = 5$ will show two peaks. Compared to the pure case (Fig. 1(b)), there are additional low-temperature anomalies around the integer values of the field, resulting from the step-wise splitting of the magnetization curves.

If we continue diluting the system to the concentration $p = 0.9$, we can observe further interesting evolution of the magnetization and susceptibility dependencies. Namely, the magnetization dependence in Fig. 3(a) indicates (and the sublattice magnetizations confirm) that the ferrimagnetically ordered region shrinks again and is now limited to the field range $1 < h/|J| < 5$. The reentrant behavior observed at p = 0.9 does not show any more, however, a different kind of the reentrant behavior appears. In particular, by fixing the temperature just above $k_BT/|J| = 0.4$ and varying the field, the system can enter the ferrimagnetic phase twice. Thus the curve of the transition related susceptibility maxima is again deformed, as shown in Fig. 3(b), and the step-wise magnetization related maxima get more prominent.

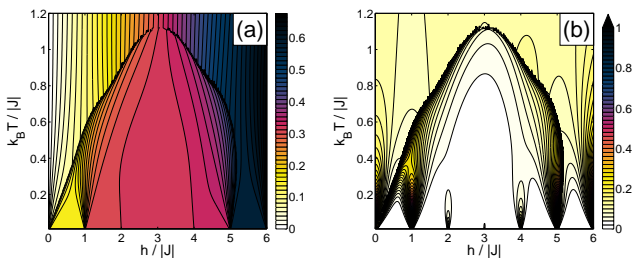

Fig.2. The same quantities as in Fig. 1, for $p = 0.95$.

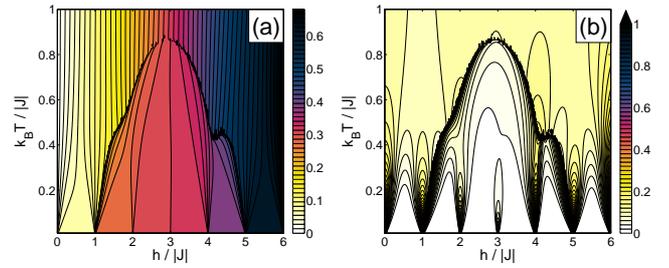

Fig.3. The same quantities as in Fig. 1, for $p = 0.9$.

## Conclusions

We studied the geometrically frustrated triangular Ising antiferromagnet, focusing on effects of an external magnetic field and random site dilution on the behavior of some thermodynamic quantities, such as the magnetization and the magnetic susceptibility. It was found that the interplay of the geometrical frustration, magnetic dilution, thermal fluctuations and the applied magnetic field may lead to different anomalies in their behavior. We presented these quantities as continuous functions in the temperature-field plane and their evolution with progressing dilution.

## Acknowledgement

This work was supported by the Scientific Grant Agency of Ministry of Education of Slovak Republic (Grant No. 1/0234/12). The authors acknowledge the financial support by the ERDF EU (European Union European Regional Development Fund) grant provided under the contract No. ITMS26220120047 (activity 3.2.).